\documentclass[sigconf]{acmart}
\AtBeginDocument{%
  }

\usepackage{todonotes}
\begin{document}

\title{Lazy Arithmetic using Systolic Arrays for \\ Closing the Verification Gap on Embedded Systems}

\author{Taisa Kushner}
\email{taisa@galois.com}
\affiliation{%
  \institution{Galois Inc}
  \city{Boulder}
  \state{Colorado}
  \country{USA}
}

\author{Ryan McCleeary}
\affiliation{%
  \institution{Galois Inc}
  \city{Fort Collins}
  \country{Colorado}
  \country{USA}}
\email{mccleeary@gmail.com}

\author{Martin Nyx Brain}
\affiliation{%
  \institution{City St George's, Univ. of London}
  \city{London}
  \country{England}
  \country{UK}}
\email{martin.brain@citysteorges.ac.uk}

\maketitle

\section{Overview}
As capabilities of deep neural network (DNN) models has grown -- from computer vision systems to digital twins driving precision medicine -- so has the interest in verifying these models for deployment in safety-critical embedded devices. However, the deployment of these models, even formally verified ones, has laid bare a key limitation: current software (SW) and hardware (HW) approaches for deploying edge DNNs are designed for increased throughput, not safety, correctness, nor fault-tolerance. As such, guarantees of model correctness are breaking. 


\section{Limitations of Current Verification Schemes}
While a range of DNN verification schemes exist, they are largely perform model verification on ``evaluation HW'' (computers, servers, etc)~\cite{george2026torchlean, Narasimhamurthy2019VerifyingCO, betaCrown, NeuralSAT2025}. This presents challenges for embedded and edge device deployment due to numerical incompatibilities across HW platforms. For example, in \cite{10.1007/978-3-030-88806-0_9} the authors show that floating point errors can be exploited in the verification of DNNs, negating verification guarantees. While they lay out ways to mitigate this issue the proposed solutions (including large over-approximates, relying on specific details of specific floating point base types, using heavily quantized for fixed point or even binary NN, and/or using exact real arithmetic) all such solutions are currently far too inefficient for practical applications. 
This results in a breaking of formally verified guarantees when porting models between HW platforms.

\section{Deployment Breaks: Software Quantization}
As full-precision inference of DNNs with acceptable latency and memory use is often infeasible on edge devices, a common remedy is quantization. This is the process of representing weights and activations with reduced and finite-precision arithmetic to decrease compute and storage costs. State-of-the-art practice includes quantization-aware training as well as post-training quantization (PTQ); more recently, automated mixed-precision PTQ methods have shown that per-layer bit-widths can be selected to reduce inference cost while providing provable error bounds under fixed modeling and operating assumptions ~\cite{Lohar_2023}. However, existing approaches largely assume static bit-width assignments, fixed input distributions/classes, and unchanging performance bounds, which limits applicability in real deployments where operating conditions and required accuracy can vary with input range, and operating environment. In addition, we lack a principled framework to identify and reason about inputs that are most sensitive to quantization-induced shifts, particularly those near decision boundaries where small numeric perturbations can change predicted classes and void formal verification guarantees.

\section{Deployment Breaks: Edge Hardware}
Even assuming a verified, sound, and efficient quantization scheme existed, current edge hardware such as GPUs, NPUs, and TPUs prioritize throughput and computational efficiency, rather than robustness against low-level hardware attacks. In particular, recent work has shown significant risk of embedded DNNs exhibiting ``silent degradation'' due to bit flips, either through fault-injection (FI) attacks~\cite{Leveugle_2025} or benign single-event upsets (SEUs) from environmental radiation~\cite{SEUs-Embedded-Gutierrez2024}. This breaks any formal guarantees of DNN model correctness as evaluated on a development platform. 

\section{Our Work and Proposal}
As such, novel efficient and soundly verified quantization schemes as well as fault tolerant devices performing evaluations of neural networks on the edge are growing in need. Herein we present two interwoven ideas: a SW approach for enabling real-time adaptive-precision quantization, along with the HW to support it. The quantization algorithm dynamically adjusts precision online while performing sensitivity analysis to quantify and manage the risk of decision-boundary crossings, enabling resource-efficient inference with explicit, verified, reasoning about quantization impacts. To achieve this we utilize a lazy left-to-right approach leveraging lazy multi-word floats, continued fractions and continued logarithms. This approach is not limited to DNNs, though we show how the specific architecture of DNNs can be leveraged for significant efficiency gains, passing the most-significant blocks (MSB) -- whether these are bits, floats, or other types -- quickly, and removing unnecessary computations in real-time. Next we propose how systolic arrays can be designed to enable even stronger efficiency of this new quantization scheme, and use triplication to ensure resilience to bit flip attacks on the MSB \cite{10.1145/3787501.3787503}. 

Together, these advances can close the gap between existing formal verification techniques for DNN algorithms, and their verified, space and memory efficient deployment onto embedded devices. This submission is presented as work-in-progress, with SW implementations completed and HW in-progress.

\section{Verified and Sound Dynamic, Adaptive, Real-Time Quantization (DARQ)}
In a recent project, we showed three key contributions, detailed below which provide the basis for our quantization scheme, and systolic hardware idea. Furthermore, we constructed a DNN evaluation framework which can both identify when and how various quantization schemes will affect DNN performance, as well as provide a sound left-to-right implementation for DNNs. 
\paragraph{Quantization Error is Rare but Significant} By evaluating a suite of image classification DNNs, we have shown that when and where quantization error effects models is rare, difficult to predict, but very significant. Existing results include a suite of models trained on fuzzy-boundary Fashion-MNIST dataset~\cite{fashion-mnist}, with others in progress. This aligns with current standing that intensive quantization is generally OK, though in the case of DNNs where correctness is of the essence, such as safety-critical systems, errors can be severely detrimental. Furthermore, determining when and where these errors arise is non-trivial, pointing to a unique benefit of our lazy arithmetic approach in identifying these points in real-time. Importantly, we are able to perform this across all operations within the DNN, not only at the output layer. 
\paragraph{Lazy Computation is Possible} We show it is possible (in theory and practice) to utilize lazy `left-to-right'' arithmetic to enable sound, power and memory efficient computation. We have shown this potential in enabling the use of neural network based algorithms in complex environments and on resource-limited hardware where performance guarantees are necessary. This lays the groundwork for implementation on both existing and custom hardware, as with the systolic arrays. 
\paragraph{Leveraging Neural Network Architectures for Significant Efficiency Gains:} While this is broadly useful for high assurance mathematics, graphical applications and signal processing, the architecture of DNN models is such that significant power and memory efficiency gains are possible through our lazy implementation using a lazy list of small floats: we find locations when upwards of 99\% of computations are unused, and we can stop these operations quickly through lazy implementations, or target them for probabilistic optimization. An example of one such DNN evaluated is provided in Table~\ref{table:depth:cnn:all}, where it is evident that certain layer types (such as Max Pool) require both an average and worst-case high number of MSBs to achieve sound evaluation (mostly caused by true equalities of two separate but complex chain of arithmetic), whereas other layer types such as Convolutions with ReLu`s never require more than a depth of 2.

\begin{table}
    \centering
    \begin{tabular}{lccccc}
    Layer Type & Avg & Med & Max & Depth $>1$ & Comparisons \\
    Convolution & 1.004 & 1 & 2 & 0.4\% & 5290000 \\
    Max Pool & 1.229 & 1 & 1151 & 18.3\% & 3630000 \\
    Full Network & 1.095 & 1 & 1151 & 7.7\% & 8920000 \\
    \end{tabular}
    \caption{Depth Analysis of MSBs (utilizing a base FP8E5M2) required for an example DNN evaluation, grouped by layer type. Large max depths were true equalities.}
    \label{table:depth:cnn:all}
\end{table}

\section{Leveraging DNN Architecture for Systolic Array Hardware Efficiencies}
Building from the final key contribution of our software quantization approach, we further define the particular aspects of DNNs which enable both software and hardware efficiency. The foundational arithmetic used in DNNs relies on large matrix math operations followed by a non-linear activation function. These models are designed to be sparse, and as such often times these activation functions can be answered with very little information from the calculation just performed. For example the rectified linear unit (ReLU) activation function simply checks if the function is positive or negative. This information could be answered by simply checking the sign of the calculation performed. In our noted DARQ work, we found that often a very small amount of information is actually needed for DNNs to evaluate -- at times with upwards of 99\% of computations able to be stopped after only 1 MSB (Table~\ref{table:depth:cnn:all}).  

While we can currently evaluate DNNs using lazy arithmetic to calculate MSBs, there is significant overhead associated with these lazy computations done in software. 
Thus we turn to novel HW and are currently working on lazy arithmetic system using systolic arrays in which the MSBs (with a bit-base) are calculated first. This allows for calculating activation functions quickly and efficiently, while also identifying the critical bits necessary to flow through a DNN to boost resilience to bit flip attacks. These bits can be queued and pulled from for other calculations further down the DNN. In order to accomplish this task we build upon the algorithms laid out in \cite{10.5555/646539.696030}, and detailed below. We have chosen the algorithms in \cite{10.5555/646539.696030} as a starting point due to it's simplicity and for the fact that the algorithms allow for the ability to get down to MSBs with bit bases (rather than floats or other types).

\section{Real Number Algorithms for Systolic Arrays}
\begin{figure}[h]
  \centering
  \includegraphics[width=\linewidth]{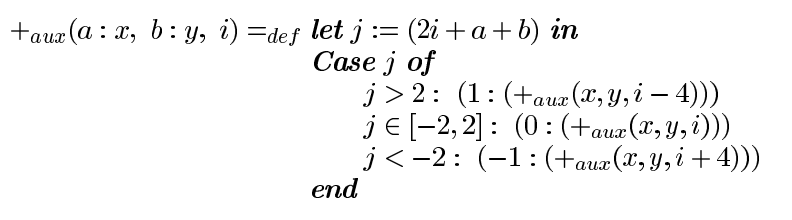}
  \caption{Addition algorithm for real number streams \cite{10.5555/646539.696030}. } 
\label{fig:addition}
\end{figure}

In \cite{10.5555/646539.696030}, a lazy stream of ternary digits is used to represent the real numbers. Addition of these streams is done by consuming one ternary digit from each stream, to create an initial state, and after that the algorithm produces the most significant digit by consuming one more from each, Figure~\ref{fig:addition}. Multiplication of scalars is similarly defined in \cite{10.5555/646539.696030}.

Currently we are generating a systolic implementation of this algorithm that breaks down two doubles into this format and produces the most significant digit first. It does this by creating a queue of digits from each double, and then pumps them through the gate to get a single queue output, which has the most significant bit first. 

We are also exploring chaining these together to form general matrix multiply (GeMM) algorithms and multiply add (MAD) algorithms commonly used in DNN, which produces a queue of outputs with the MSB first. The eventual end goal being a library of algorithms similar to that of \cite{cuDNN}.

\section{Significance and Conclusion}
Together the SW and HW approaches presented herein fill a significant gap in verifying DNNs for use on embedded safety-critical devices, by enhancing efficiency, soundness, and bridging the gap between reasoning with floating point numbers and reals. 
If our work proves fruitful, it would allow reasoning directly with a real number implementation to allow for an easier verification story, and mitigate issues between algorithm verification and implementation on edge devices. Furthermore, the novel implementation of the real numbers using systolic arrays which we are exploring is not only limited to DNNs, but generalizes across all high-precision mathematics needs. Lastly, the MSB targeted approach allows for robustness against bit flip attacks (malicious or spurious) which is of interest for edge devices.

\section*{Acknowledgments}
This material is based upon work supported by the Defense Advanced Research Projects Agency (DARPA) and Naval Information Warfare Center Pacific (NIWC Pacific) under Contract
No. N66001-21-C-4023. Any opinions, findings and conclusions or recommendations expressed in this material are those of the author(s) and do not necessarily reflect the views of
DARPA and NIWC Pacific. Distribution Statement A. Approved for public release: distribution is unlimited.


\bibliographystyle{ACM-Reference-Format}
\bibliography{array}


\end{document}